\documentclass[10pt,final,conference,letterpaper]{IEEEtran}
\usepackage[english]{babel}
\usepackage[latin1]{inputenc}
\usepackage{mathenv}
\usepackage{amsmath}
\usepackage{amsfonts}
\usepackage{epsfig}
\usepackage{units}
\usepackage{nicefrac}
\usepackage{pstricks}
\usepackage{pst-node}
\usepackage{pstricks-add}

\newcommand{\Prob}[0]{\text{Pr}}

\newcommand{\etal}[0]{\textit{et al.}}
\newcommand{\CF}[0]{\text{CF}}
\newcommand{\DF}[0]{\text{DF}}
\newcommand{\Set}[1]{\mathcal{#1}}
\newcommand{\RV}[1]{\mathrm{#1}}
\newcommand{\Markov}[0]{\leftrightarrow}
\newcommand{\I}[0]{\mathrm{I}\!}
\newcommand{\Vector}[1]{\mathbf{#1}}

\newcommand{\StronglyTypicalSet}[0]{\Set{A}_\epsilon^{*(n)}}

\def\isequalto{:=}

\newtheorem{theorem}{Theorem}

\newtheorem{corollary}{Corollary}
\newtheorem{definition}{Definition}

\title{Protocols For Half-Duplex Multiple Relay Networks}
\author{
\authorblockN{Peter Rost, \textit{Student Member, IEEE}, and Gerhard Fettweis, \textit{Senior Member, IEEE}}
  \authorblockA{
    Technische Universität Dresden, Vodafone Chair Mobile Communications Systems, Dresden, Germany\\
    EMail: \{rost, fettweis\}@ifn.et.tu-dresden.de
  }
}
\IEEEoverridecommandlockouts
\begin{document}
\maketitle
  \begin{abstract}
    In this paper we present several strategies for multiple relay networks which are
    constrained by a half-duplex operation, i.\,e., each node either transmits or
    receives on a particular resource. Using the discrete memoryless multiple relay
    channel we present achievable rates for a multilevel partial decode-and-forward
    approach which generalizes previous results presented by Kramer and Khojastepour \etal.
    Furthermore, we derive a compress-and-forward approach using a regular encoding
    scheme which simplifies the encoding and decoding scheme and improves
    the achievable rates in general. Finally, we give achievable rates for a mixed strategy used
    in a four-terminal network with alternately transmitting relay nodes.
  \end{abstract}
  \section{Introduction}
    Infrastructure based wireless communications systems as well as ad hoc networks form an integral part
    of our everyday life. An increased density and availability of mobile terminals pose
    the question which techniques next generation networks shall employ to improve reliability and data rate.
    One way to exploit the capabilities of these networks is the use of \emph{relay nodes} which support communication pairs.
    The idea of relaying was introduced in \cite{vdMeulen.TR.1968} and
    substantially refined for the three-terminal case in \cite{Cover.Gamal.TransIT.1979}.
    
    More recent publications focus their attention on relay networks of arbitrary size, e.\,g.,
    \cite{Kramer.Gastpar.Gupta.TransIT.2005} presents general coding strategies using 
    different \emph{decode-and-forward} (DF) and \emph{compress-and-forward} (CF) approaches. When relay nodes are
    cooperating using decode-and-forward, they must decode the complete source message and provide
    additional information similar to Slepian-Wolf coding \cite{Slepian.Wolf.1973}. In contrast, when following
    a compress-and-forward approach, each relay quantizes its own channel output which has to
    be decoded by the actual information sink (similar to Wyner-Ziv coding \cite{Wyner.Ziv.1976}).

    Practical restrictions as well as cost issues imply an \emph{orthogonality constraint} on relay nodes, i.\,e.,
    in contrast to the previously mentioned work we consider
    half-duplex terminals which either transmit or listen on a particular resource. 
    First information-theoretical results considering this constraint were presented for the three-terminal network
    in \cite{Khojastepour.Sabharwal.Aazhang.Globecom.2003,Madsen.VTC.2002}.
    For the $N$-terminal case, \cite{Khojastepour.Sabharwal.Aazhang.IPSA.2003} derives upper bounds on the achievable rates.
    While these papers assume fixed transmission schedules known to all nodes,
    a new strategy was presented in \cite{Kramer.2004} for the three-terminal case where the node states, i.\,e.,
    sleep, listen or transmit, are used to exchange information.

    In the sequel we will take up the idea of \cite{Kramer.2004} and present more general formulations for relay networks
    of arbitrary size.  First, we introduce in Section \ref{section:system.model} the channel model for the half-duplex
    multiple relay network. Afterwards, we discuss in Section \ref{sec:protocol.structure:df}
    a partial decode-and-forward protocol based on the regular encoding approach introduced in \cite{Xie.Kumar.TransIT.2005}.
    Then, we present in Section \ref{sec:protocol.structure:cf} a generalized
    compress-and-forward approach using a regular encoding structure which might be of interest
    for other problems such as the successive refinement problem \cite{Equitz.Cover.TransIT.1991}.
    Finally, we derive a mixed protocol for two alternately transmitting relay nodes in Section \ref{sec:protocol.structure:mixed}.
    This scheme is dedicated to an application in wireless networks where each relay has only sufficient
    channel conditions either to the source or destination. 

  \section{Network model, nomenclature and definitions}\label{section:system.model}
    In the following we will use non-italic uppercase letters
    $\RV{X}$ to denote random variables, non-italic lowercase letters $\RV{x}$ to denote
    events of a random variable (r.v.) and italic letters ($N$ or $n$) to denote constant values.
    Ordered sets are denoted by $\Set{X}$, the cardinality of an ordered set is denoted by $\left\|\Set{X}\right\|$
    and $\left[b; b+k\right]$ is used to denote the ordered set of numbers $\left(b, b+1, \cdots, b+k\right)$.
    Let $\RV{X}_k$ be a random variable parameterized by $k$, then 
    $\RV{\Vector{X}}_{\Set{C}}$ denotes the vector and $\left\{\RV{X}_k\right\}_{k\in\Set{C}}$ the set
    of all $\RV{X}_k$ with $k\in\Set{C}$ (this applies similarly to sets of events). 
    Furthermore, we will use $p(\RV{x} | \RV{y})$ to abbreviate the conditional probability density function (pdf)
    $p_{\RV{X} | \RV{Y}}(\RV{x} | \RV{y})$ for the benefit of readability. 
    $\I\left(\RV{X}; \RV{Y} | \RV{Z}\right)$ denotes the mutual information between r.v.s $\RV{X}$ and $\RV{Y}$
    given $\RV{Z}$ \cite{Cover.Thomas.1991}
    
    This paper considers a network of $N+2$ nodes: the source node $s=0$, the set of $N$ relays $t\in\Set{R}\isequalto[1;N]$
    and the destination node $d=N+1$.
    The discrete memoryless multiple relay channel is defined by the conditional pdf
    $p\left(\RV{\Vector{y}}_{[1;N+1]} | \RV{\Vector{x}}_{[0;N]}, \RV{\Vector{m}}_{[0;N]}\right)$
    over all possible channel inputs
    $\left(\RV{x}_s, \RV{x}_1, \cdots, \RV{x}_{N}\right)\in\Set{X}_s\times\Set{X}_1\times\cdots\Set{X}_{N}$,
    channel outputs $\left(\RV{y}_1, \cdots, \RV{y}_{N}, \RV{y}_{d}\right)\in\Set{Y}_1\times\cdots\Set{Y}_{N}\times\Set{Y}_{d}$
    and node states $\left(\RV{m}_s, \RV{m}_1, \dots, \RV{m}_N\right)\in\Set{M}_s\times\Set{M}_1\times\dots\Set{M}_N$ with 
    $\Set{M}_t=\left\{L, T\right\}$. Each $t\in[0;N]$
    is either listening ($\RV{M}_t=L$) or transmitting ($\RV{M}_t=T$) on a particular resource. 
    In contrast to \cite{Kramer.2004} we do not consider a possible sleep state where the node is
    neither listening nor transmitting.  Besides,
    it is possible that the source remains silent, e.\,g., to reduce interference in a wireless network.
    As an immediate consequence of the orthogonality constraint
    we can state that $\left(\RV{M}_t=T\right)\rightarrow\left(\RV{Y}_t=\varphi\right)$ and $\left(\RV{M}_t=L\right)\rightarrow\left(\RV{X}_t=\psi\right)$
    where $\varphi$ and $\psi$ are arbitrary, known constants. The previous definitions further assume that the destination is always listening.
    
    Let $\pi(\Set{X})$ be the set of all permutations of a set $\Set{X}$.
    The source chooses an ordering $o_s\in\pi([1;N+1])$ where
    $o_s(l)$ denotes the $l$-th element of $o_s$ and $o_s(N+1)=N+1$. For the sake of readability,
    we abbreviate in the following $\RV{Y}_{o_s(l)}$ by $\RV{Y}_l$ and 
    the relay node $o_s(l)$ by $l$ or as the $l$\emph{-th level}. 
    All results presented in the sequel are given for a specific $o_s$, though a
    maximization over $\pi([1;N+1])$ is necessary.

    We further divide all transmissions in blocks $b\in[1;B]$ of length $n$. Now, consider the
    following standard definitions:
    \begin{definition}
      A $(2^{nR}, n, \lambda_n)$ code for the multiple relay channel consists of
      \begin{itemize}
	\item a set of indices $\Set{W}=[1; 2^{nR}]$ with equal probability and the corresponding
	  r.v. $\RV{W}$ over $\Set{W}$,
	\item the source encoding function $f_0: [1; 2^{nR}]\rightarrow \Set{X}_s^n \times \Set{M}_s^n$,
	\item relay encoding functions $f_{l; b}: \Set{Y}_l^{n\cdot(b-1)}\rightarrow \Set{X}_l^n \times \Set{M}_l^n$,
	\item the decoding function $g: \Set{Y}_d^n\rightarrow [1; 2^{nR}]$,
	\item and the maximum probability of error 
	  \begin{equation*}
	    \lambda_n = \max\limits_{w\in\Set{W}}\Prob\left\{g(\RV{y}_d)\neq w | \RV{W}=w\right\}.
	  \end{equation*}
      \end{itemize}
    \end{definition}

    \begin{definition}
      A rate $R$ is achievable if there exists a sequence of
      $(2^{nR}, n, \lambda_n)$ codes such that $\lambda_n\rightarrow0$ as $n,B\rightarrow\infty$.
    \end{definition}

  \section{Decode-and-forward protocols}\label{sec:protocol.structure:df}
    The first protocols we present in this paper are an application of the partial decode-and-forward
    approach \cite{Cover.Gamal.TransIT.1979} to multiterminal half-duplex relay networks. 
  
    \subsection{Multilevel partial decode-and-forward}\label{sec:protocol.structure:df:pdf}
      \begin{figure}
	\centering
	\begingroup
\unitlength=1mm
\begin{picture}(76, 44)(0, 0)
  \psset{xunit=1mm, yunit=1mm, linewidth=0.2mm}
  \rput(2, 10){\cnodeput(0, 0){Source}{$s$}}
  \rput(25, 35){%
    \cnodeput(0, 0){R1}{$1$}
    \rput(0, 7){$\RV{Y}_1 : \RV{U}_s^1$}}%
  \ncarc{->}{Source}{R1}\mput*{$\RV{U}_s^1$}
  \rput(55, 35){%
    \cnodeput(0, 0){R2}{$2$}
    \rput(0, 7){$\RV{Y}_2 : \RV{U}_s^1\RV{U}_s^2$}}%
  \ncarc{->}{R1}{R2}\mput*{$\RV{V}_1^1$}
  \ncarc{->}{Source}{R2}\mput*{$\RV{U}_s^1\RV{U}_s^2$}
  \rput(65, 10){\cnodeput(0, 0){Destination}{$d$}\rput(0, -6){$\RV{Y}_d:\RV{U}_s^1\RV{U}_s^2\RV{U}_s^3$}}
  \ncarc{->}{Source}{Destination}\mput*{$\RV{U}_s^1\RV{U}_s^2\RV{U}_s^3$}
  \ncarc{->}{R1}{Destination}\mput*{$\RV{V}_1^1$}
  \ncarc{->}{R2}{Destination}\mput*{$\RV{V}_2^1\RV{V}_2^2$}
\end{picture}
\endgroup
	\caption{\emph{Information exchange} of partial decode-and-forward for $N=2$.}
	\label{fig:protocol:information.exchange}
      \end{figure}
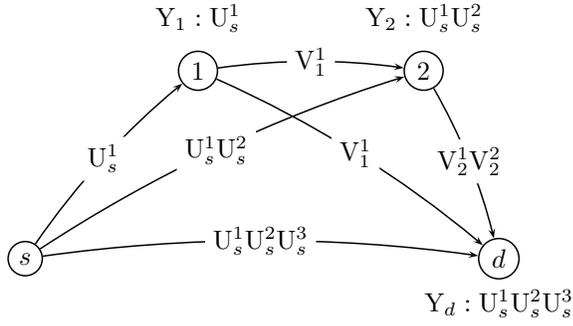
      Our first proposal is a partial decode-and-forward approach illustrated in Fig. \ref{fig:protocol:information.exchange}.
      The source message $\RV{W}$
      is mapped to the tuple $\left(\RV{M}_s, \RV{U}_s^1, \dots, \RV{U}_s^{N+1}\right)$, with $\RV{U}_s^k\in[1; 2^{nR_s^k}]$. 
      As previously mentioned, we have a specific ordering $o_s$ which implies that each relay $l\in[1;N+1]$ must decode the source messages
      $\RV{U}_s^{[1;l]}$ and provides additional information by transmitting the independently
      generated message tuple $\left(\RV{M}_l, \RV{V}_l^1, \dots, \RV{V}_l^l\right)$, with $\RV{V}_l^k\in[1; 2^{nR_s^k}]$. 
      Using the example in Fig. \ref{fig:protocol:information.exchange}, relay $1$ decodes $\RV{U}_s^1$ and transmits
      the support message $\RV{V}_1^1$, whereas relay $2$ decodes the tuple $\left(\RV{U}_s^1, \RV{U}_s^2\right)$ and
      provides additional information with the tuple $\left(\RV{V}_2^1, \RV{V}_2^2\right)$.
      Relay $2$ can additionally exploit $\RV{V}_1^1$ to decode $\RV{U}_s^1$.
      As we employ a Markov superposition coding, node $l$ transmits in block $b$ additional information
      for the source messages transmitted in block $b-l$. In this way, we ensure that level $l$ is able
      to support the transmission of relays $l'>l$ and message levels $[1;l]$. 
      
      Finally, when decoding the source message level $k\in[1;N+1]$ in block $b$, the destination jointly decodes the
      messages $\RV{U}_s^k$ transmitted in block $b-N$, and the relay message $\RV{V}_l^k$ transmitted in block $b-l+1$ for $l\in[k;N]$.
      Furthermore, if $k=1$ the destination also decodes the node states $\RV{M}_{[0;N]}$ which carry
      additional information.
      This regular encoding and decoding structure was introduced in \cite{Xie.Kumar.TransIT.2005} and
      applied to a mixed protocol structure for full-duplex networks in \cite{Rost.Fettweis.TransIT.2007}.
      Now we are able to formulate the following theorem:
      \begin{theorem}\label{theorem:half.duplex.random:1}
	The achievable rate $R=\sum\limits_{k=1}^{N+1}R_s^k$ using partial
	decode-and-forward with a random  schedule is given by
	\begin{equation}
	  \begin{split}
	    R_s^1 & \leq \sup\limits_p\min\limits_{l\in[1;N+1]}
	    \I\left(\RV{M}_s,\RV{U}_s^1; \RV{Y}_{l} | \left\{\RV{V}_{[i;N]}^i\right\}_{i=1}^l, \RV{M}_{[1;N]}\right) \\
	    &\quad{+}\;\sum\limits_{j=1}^{l-1} \I\left(\RV{M}_j,\RV{V}_j^1; \RV{Y}_{l} | \left\{\RV{V}_i^{[1;i]}\right\}_{i=j+1}^l, \RV{V}_{[l;N]}^{[1;l]}, \RV{M}_{[j+1;N]}\right)
	   \end{split}
	   \label{eq:df:pdf:10}
	\end{equation}
	\begin{equation}
	  \begin{split}
	    R_s^k & \leq \sup\limits_p\min\limits_{l\in[k;N+1]}
	    \I\left(\RV{U}_s^k; \RV{Y}_{l} | \RV{U}_s^{[1;k-1]}, \left\{\RV{V}_{[i;N]}^i\right\}_{i=1}^l, \RV{M}_{[0;N]}\right) \\
	    & \quad {+} \sum\limits_{j=k}^{l-1} \I\left(\RV{V}_j^k; \RV{Y}_{l} | \RV{V}_j^{[1;k-1]}, \left\{\RV{V}_i^{[1;i]}\right\}_{i=j+1}^l, 
	    \RV{V}_{[l;N]}^{[1;l]}, \RV{M}_{[j;N]}\right)
	  \end{split}
	  \label{eq:df:pdf:11}
	\end{equation}
	for $k\in[2;N+1]$. The supremum in (\ref{eq:df:pdf:10}) and (\ref{eq:df:pdf:11}) is taken over all joint pdfs of the form
	\begin{equation}
	  \begin{split}
	    & p\left(\RV{y}_{[1;N+1]}, \RV{u}_s^{[1;N+1]}, \RV{v}_{l\in[1;N]}^{[1;l]}, \RV{m}_{[0;N]}\right) = \\
	    & \quad p\left(\RV{y}_{[1;N+1]} | \RV{u}_s^{[1;N+1]}, \RV{v}_{l\in[1;N]}^{[1;l]}\right)\cdot\prod\limits_{l=s}^N p\left(\RV{m}_l | \RV{m}_{[l+1;N]}\right)\\
	    & \quad {\cdot}\;\prod\limits_{l=1}^N\prod\limits_{k=1}^l p\left(\RV{v}_l^k | \RV{v}_l^{[1;k-1]}, \RV{v}_{[l+1;N]}^k, \RV{m}_{[l;N]}\right)\\
	      & \quad {\cdot}\;\prod\limits_{k=1}^{N+1}p\left(\RV{u}_s^k | \RV{u}_s^{[1;k-1]}, \RV{v}_{l\in[k;N]}^{k}, \RV{m}_{\left\{s,[k;N]\right\}}\right).
	  \end{split}
	  \label{eq:df:pdf:12}
	\end{equation}
      \end{theorem}
      \begin{proof}
	Using the result given in \cite[Theorem 1]{Rost.Fettweis.TransIT.2007} we apply the substitutions 
	$\RV{U}_s^1\mapsto\left(\RV{U}_s^1, \RV{M}_s\right)$ and $\RV{V}_l^1\mapsto\left(\RV{V}_l^1, \RV{M}_l\right)$ and skip
	the CF part, yielding the joint pdf in (\ref{eq:df:pdf:12}). 
	Eq. (\ref{eq:df:pdf:10}) can be slightly simplified by modifying (\ref{eq:df:pdf:12}) such that
	the Markov condition $\RV{M}_s\Markov\RV{U}_s^1\Markov U_s^{[2;N+1]}$ is satisfied (and similar for all relay messages) which
	yields the results given in \cite{Kramer.2004}.
      \end{proof}
      
      In the previous theorem we assumed a random channel access by each node. 
      To improve for instance the interference mitigation in wireless networks it might be preferable to have a fixed
      transmission scheme (beside the fact that the random access strategy can provide at most an improvement of $N+1$ bits).
      Therefore, consider the following corollary:
      \begin{corollary}[to Theorem \ref{theorem:half.duplex.random:1}]\label{corollary:half.duplex.random:1}
	In case of a fixed strategy known to all nodes, we can achieve any rates satisfying
	\begin{equation*}
	  \begin{split}
	    R_s^k & \leq \sup\limits_p\min\limits_{l\in[k;N+1]}
	    \I\left(\RV{U}_s^k; \RV{Y}_{l} | \RV{U}_s^{[1;k-1]}, \left\{\RV{V}_{[i;N]}^i\right\}_{i=1}^l, \RV{M}_{[0;N]}\right) \\
	    & \quad {+} \sum\limits_{j=k}^{l-1} \I\left(\RV{V}_j^k; \RV{Y}_{l} | \RV{V}_j^{[1;k-1]}, \left\{\RV{V}_i^{[1;i]}\right\}_{i=j+1}^l, 
	    \RV{V}_{[l;N]}^{[1;l]}, \RV{M}_{[0;N]}\right)
	  \end{split}
	\end{equation*}
	for all $k\in[1;N+1]$. The supremum is taken over all joint pdfs similar to (\ref{eq:df:pdf:12}) with the appropriate
	changes reflecting that $\RV{M}_{[0;N]}$ is now known to all nodes.
      \end{corollary}

    \subsection{Multilevel decode-and-forward}\label{sec:protocol.structure:df:df}
      Assume that the source uses only a single message level. In this case
      we obtain an application of the multilevel DF protocol presented in \cite{Xie.Kumar.TransIT.2005} to
      half-duplex networks. The achievable rates are summarized in the following corollary:
      \begin{corollary}[to Theorem \ref{theorem:half.duplex.random:1}]\label{corollary:half.duplex.random:3}
	The achievable rate $R$ using multilevel DF with a random schedule is given by
	\begin{equation}
	  R \leq \sup\limits_p\min\limits_{l\in[1;N+1]} \I\left(\RV{M}_{[0;l-1]}, \RV{X}_{[0;l-1]}; \RV{Y}_{l} | \RV{X}_{[l;N]}, \RV{M}_{[l;N]}\right).
	  \label{eq:df:multilevel:10}
	\end{equation}
	For fixed transmission strategies it is given by
	\begin{equation}
	  R \leq \sup\limits_p\min\limits_{l\in[1;N+1]} \I\left(\RV{X}_{[0;l-1]}; \RV{Y}_{l} | \RV{X}_{[l;N]}, \RV{M}_{[0;N]}\right).
	  \label{eq:df:multilevel:11}
	\end{equation}
	In both cases the supremum is taken over all joint pdfs of the form given in (\ref{eq:df:pdf:12})
	with $k=1$ instead of $k\in[1;N]$.
      \end{corollary}
      It turns out that the rates are in general lower than in Theorem \ref{theorem:half.duplex.random:1}. Besides,
      it shows that the previously described protocols generalize the results presented in
      \cite{Khojastepour.Sabharwal.Aazhang.Globecom.2003} and \cite{Kramer.2004}.

    \subsection{Multihopping with limited resource reuse}\label{sec:protocol.structure:df:multihopping}
      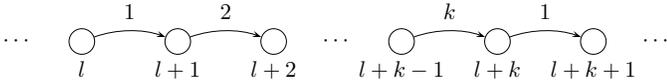
\begin{figure}
	\centering
	\scalebox{0.85}{\begingroup
\unitlength=1mm
\begin{picture}(104, 13)(0, 0)
  \psset{xunit=1mm, yunit=1mm, linewidth=0.1mm}
  \psset{arrowsize=2pt 3, arrowlength=1.4, arrowinset=.4}

  \rput(0, 7){$\cdots$}
  \rput(10, 7){\cnode(0,0){2mm}{L1}}
  \rput[t](10, 4){$l$}
  \rput(25, 7){\cnode(0,0){2mm}{L2}}
  \rput[t](25, 4){$l+1$}
  \ncarc[arcangle=20, linestyle=solid]{->}{L1}{L2}\naput{$1$}
  \rput(40, 7){\cnode(0,0){2mm}{L3}}
  \rput[t](40, 4){$l+2$}
  \ncarc[arcangle=20, linestyle=solid]{->}{L2}{L3}\naput{$2$}

  \rput(50, 7){$\cdots$}

  \rput(60, 7){\cnode(0,0){2mm}{L4}}
  \rput[t](60, 4){$l+k-1$}
  \rput(75, 7){\cnode(0,0){2mm}{L5}}
  \rput[t](75, 4){$l+k$}
  \ncarc[arcangle=20, linestyle=solid]{->}{L4}{L5}\naput{$k$}
  \rput(90, 7){\cnode(0,0){2mm}{L6}}
  \rput[t](90, 4){$l+k+1$}
  \ncarc[arcangle=20, linestyle=solid]{->}{L5}{L6}\naput{$1$}
  \rput(100, 7){$\cdots$}
\end{picture}
\endgroup}
	\caption{Multihopping with limited resource reuse $\nicefrac{1}{k}$. Edge labeling identifies the used 
	resource for the respective transmission.}
	\label{figure:halfduplex.random:multihopping}
      \end{figure}
      This case treats multihopping protocol with limited resource reuse as discussed in \cite{Herhold.Zimmermann.Fettweis.JCN.2005}. Consider the network in
      Fig. \ref{figure:halfduplex.random:multihopping} showing an example for multihopping with reuse factor $\nicefrac{1}{k}$. This implies
      that one resource is only occupied by $\nicefrac{1}{k}$-th of all nodes, or that one node only uses $\nicefrac{1}{k}$-th of
      the available resources. Applied to our half-duplex relay network this implies that the joint pdf in (\ref{eq:df:pdf:12}) must satisfy
      \begin{equation*}
	\forall l\in[0;N]: \Prob\left(\RV{m}_{l} = T \Bigl| \exists j\in[1;k-1]: \RV{m}_{l-j} = T\right) = 0,
      \end{equation*}
      that is none of the nodes in levels $[l-k+1; l-1]$ is allowed to transmit on the same resources as node $l$.

  \section{A compress-and-forward approach}\label{sec:protocol.structure:cf}
    In the previous section we presented different decode-and-forward based approaches. These protocols
    are likely to suffer from the necessity of decoding the complete source message at \emph{each} node, which is an even more
    severe drawback in half-duplex networks.
    In this section, we discuss a compress-and-forward protocol
    which might overcome this issue. We assume a fixed transmission scheme
    implying exact knowledge at each node about the current transmission state of any other node.

    More specifically, each relay $l\in[1;N]$ creates the quantization messages $\hat{\RV{Y}}_l$ and 
    the corresponding broadcast messages $\RV{X}_l$, \emph{both} with rates $\Delta_l$. Consider the transmission in block $b$:
    node $l$ searches for a quantization vector which is jointly typical with its channel output $\RV{Y}_l$
    in block $b$. Once the node found a quantization $\hat{\RV{Y}}_l$ with index $q_{l,b+1}$ it transmits
    in block $b+1$ the broadcast message $\RV{X}_l$ assigned to the same index. 
    
    Consider the decoding process at the destination  for the quantization of relay node $N$. At first
    it searches for the set of all broadcast messages $\RV{X}_N$ which are jointly typical with $\RV{Y}_d$ in block $b$.
    Furthermore, it builds the set of all quantizations $\hat{\RV{Y}}_N$ jointly typical with
    $\RV{y}_d(b-1)$ while knowing $\RV{x}_N(q_{N,b-1})$, which was decoded in the previous block. By building the intersection
    of both sets the destination is now able to decode the quantization of relay $N$ for block $b-1$. Similarly, the destination
    proceeds to decode the quantization of all other relays $l\in[1;N-1]$ where $\hat{\Vector{\RV{Y}}}_{[l+1;N]}$ is
    used to improve the rate of $\hat{\Vector{\RV{Y}}}_{l}$.
    Based on the previous description we can formulate the following theorem:
    \begin{theorem}\label{theorem:cf}
      With the previously presented compress-and-forward scheme we achieve any rate up to
      \begin{equation}
	R \leq \sup\limits_p\I\left(\RV{X}_s; \hat{\RV{Y}}_{[1;N]}, \RV{Y}_d | \RV{X}_{[1;N]}, \RV{M}_{[0;N]}\right),\label{eq:cf:10}
      \end{equation}
      subject to
      \begin{multline}
	\forall l\in[0; N-1]: \I\left(\hat{\RV{Y}}_{N-l}; \RV{Y}_{N-l} | \RV{M}_{[0;N]}\right) \leq \\
	\I\left(\hat{\RV{Y}}_{N-l}, \RV{X}_{N-l}; \hat{\RV{Y}}_{[N-l+1; N]}, \RV{Y}_d | \RV{X}_{[N-l+1; N]}, \RV{M}_{[0;N]}\right),
	\label{eq:cf:11}
      \end{multline}
      and with the supremum over all joint pdf of the form
      \begin{equation}
	\begin{split}
	  & p\left(\RV{y}_{[1;N+1]}, \RV{x}_{[0;N]}, \hat{\RV{y}}_{[1;N]}, \RV{m}_{[0;N]}\right) =
	  p\left(\RV{y}_{[1;N+1]} | \RV{x}_{[0;N]}, \RV{m}_{[0;N]}\right) \\
	  & \quad{\cdot}\;\prod\limits_{l=1}^N p\left(\hat{\RV{y}}_l | \RV{y}_l, \RV{m}_{[0;N]}\right)\cdot p\left(\RV{x}_l | \RV{m}_{[0;N]}\right).
	\end{split}
	\label{eq:cf:12}
      \end{equation}
    \end{theorem}
    \begin{proof}
      From rate distortion theory we know \cite[Ch. 13]{Cover.Thomas.1991}
      \begin{equation}
	\Delta_l \geq \I\left(\hat{\RV{Y}}_l; \RV{Y}_l | \RV{M}_{[0;N]}\right).\label{eq:cf:proof:10}
      \end{equation}
      To decode the quantization index of node $N-l$ corresponding to the destination channel output in block $b-l-1$, the destination
      searches for a $\hat q_{N-l, b-l}$ such that
      \begin{align*}
	& \exists \hat q_{N-l,b-l}: \hat q_{N-l, b-l} = \\
	& \quad \biggl\{\tilde q_{N-l, b-l}: 
	  \Bigl(\hat{\RV{y}}_{N-l}\left(\tilde q_{N-l, b-l}\right),
	  \left\{\hat{\RV{y}}_{N-l'}\left(q_{N-l', b-l}\right)\right\}_{l'=0}^{l-1}, \\
	& \quad \left\{\RV{x}_{N-l'}\left(q_{N-l', b-l-1}\right)\right\}_{l'=0}^{l},
	  \RV{y}_d\left(b-l-1\right)\Bigr)\in\StronglyTypicalSet\biggr\} \\
	& \quad {\cap}\biggl\{\tilde q_{N-l, b-l}: 
	  \Bigl(\RV{x}_{N-l}\left(\tilde q_{N-l, b-l}\right), 
	  \left\{\hat{\RV{y}}_{N-l'}\left(q_{N-l', b-l+1}\right)\right\}_{l'=0}^{l-1}, \\
	& \quad \left\{\RV{x}_{N-l'}\left(q_{N-l', b-l}\right)\right\}_{l'=0}^{l-1},
	  \RV{y}_d\left(b-l\right)\Bigr)\in\StronglyTypicalSet\biggr\},
      \end{align*}
      where $\StronglyTypicalSet$ is the $\epsilon$-strongly typical set as defined in \cite[Ch. 13.6]{Cover.Thomas.1991}. The requirement
      of \emph{strong} typicality arises from the necessity to apply the Markov lemma \cite[Lemma 14.8.1]{Cover.Thomas.1991}
      to prove joint typicality.
      The previous equation can only be fulfilled iff (\ref{eq:cf:proof:10}) holds and
      \begin{align*}
	\begin{split}
	  \Delta_{N-l} & \leq \I\left(\hat{\RV{Y}}_{N-1}; \hat{\RV{Y}}_{[N-l+1; N]}, \RV{Y}_d | \RV{X}_{[N-l; N]}, \RV{M}_{[s;N]}\right) \\
	    & \quad{+}\;\I\left(\RV{X}_{N-l}; \hat{\RV{Y}}_{[N-l+1; N]}, \RV{Y}_d | \RV{X}_{[N-l+1; N]}, \RV{M}_{[s;N]}\right)
	  \end{split}\\
	 & \leq \I\left(\hat{\RV{Y}}_{N-l}, \RV{X}_{N-l}; \hat{\RV{Y}}_{[N-l+1; N]}, \RV{Y}_d | \RV{X}_{[N-l+1; N]}, \RV{M}_{[s,N]}\right)
      \end{align*}
      Similarly the destination decodes in block $b$ the source message transmitted in block
      $b-N$ iff (\ref{eq:cf:10}) holds. Using standard methods extensively discussed in 
      literature \cite{Cover.Thomas.1991}, (\ref{eq:cf:11}) and the proof for achievability follow.
    \end{proof}
    Due to the \emph{regular encoding},
    i.\,e., quantization and broadcast messages are generated with the same rate, we are able to alleviate
    the drawbacks of source-channel coding separation. Assume multiple descriptors and an \emph{irregular encoding}.
    In this case, the decoders are forced to decode at first the broadcast and then the quantization messages where
    the first step is a severe bottleneck. For our CF scheme
    the achieved rates are the same as the destination is the only descriptor,  but the next section presents a mixed protocol 
    combining DF and CF where regular encoding can improve the achievable rates.

  \section{A mixed protocol for two relays}\label{sec:protocol.structure:mixed}
    Finally, we present a protocol for two relay nodes which are alternately transmitting. The idea of alternately
    transmitting relays goes back to \cite{Oechtering.Sezgin.2004} and achievable rates were presented in
    \cite{Xue.Sandhu.TransIT.2007} for the Diamond network
    as well as in \cite{Rost.Fettweis.Springer.2007} where DF and CF based protocols
    are discussed.

    Consider a mobile communications system where fixed infrastructure relay nodes are deployed. 
    We design the deployment such that sufficiently good channel conditions
    between relay and base station as well as between relay and mobile can be guaranteed. In networks supporting more than two hops
    it is likely to face the situation where only one relay has an excellent connection towards the base station
    and the second relay towards the mobile terminal. In this case it is recommendable to use neither a purely decode-and-forward
    based protocol nor a purely compress-and-forward based approach. 
    The latter one would be beneficial for the downlink when mobile terminals act as relay nodes whereas the former one is preferable 
    for the uplink, or if fixed relays are used in rural areas for coverage extension.
    
    \begin{figure}
      \centering
      \begingroup
\unitlength=1mm
\begin{picture}(83, 38)(0, 0)
  \psset{xunit=1mm, yunit=1mm, linewidth=0.1mm}%
  \psset{arrowsize=3pt 4, arrowlength=1.4, arrowinset=.4}%
  \rput(0, 28)%
  {%
    \rput[r](15, 0){\parbox[l]{1.2cm}{Phase 1,\\ $[1; n_1]$:}}
    \cnodeput(20, 0){S}{$s$}
    \cnodeput(40, 0){R1}{$1$}
    \cnodeput(60, 0){R2}{$2$}
    \cnodeput(80, 0){D}{$d$}
    \ncline{->}{S}{R1}\nbput{$\RV{x}_{s,1}$}
    \ncline[linestyle=dashed]{<-}{R1}{R2}\nbput{$\RV{x}_2$}
    \ncline{->}{R2}{D}\nbput{$\RV{x}_2$}
    \ncarc[arcangle=20, linestyle=solid]{->}{S}{D}\naput{$\RV{x}_{s,1}$}
  }%
  \rput(0, 8)%
  {%
    \rput[r](15, 0){\parbox[l]{1.5cm}{Phase 2,\\ $[n_1+1; n]$:}}
    \cnodeput(20, 0){S}{$s$}
    \cnodeput(40, 0){R1}{$1$}
    \cnodeput(60, 0){R2}{$2$}
    \cnodeput(80, 0){D}{$d$}
    \ncline{->}{R1}{R2}\nbput{$\RV{x}_1$}
    \ncarc[arcangle=35, linestyle=solid]{->}{S}{D}\nbput{$\RV{x}_{s,2}$}
    \ncarc[arcangle=20, linestyle=solid]{->}{S}{R2}
    \ncarc[arcangle=-40, linestyle=solid]{->}{R1}{D}
  }%
\end{picture}
\endgroup
      \caption{Example for a half-duplex channel with two alternately transmitting relay nodes.
      The solid lines indicate actual information exchange while the dashed line indicates the probably interfering
      transmission from node $2$ to $1$. The edge labeling indicates the exchanged message.}
      \label{figure:halfduplex.random:alternately_transmitting}
    \end{figure}
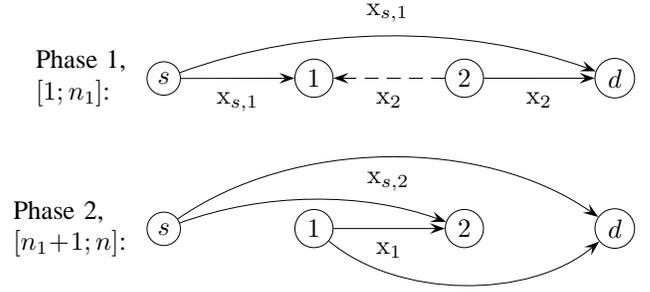
    Based on the previous motivation we will present now a protocol where one relay operates as decode-and-forward
    and the other one as compress-and-forward relay. Consider the setup illustrated in Fig. \ref{figure:halfduplex.random:alternately_transmitting}:
    the overall transmission period is divided into two phases with probabilities $p_1$ and $p_2$ such that
    \begin{align*}
      p_{\RV{M}_s}\left(T\right) & = 1, &
      p_{\RV{M}_1 | \RV{M}_2}\left(T | L\right) & = 1, \\
      p_1 & = p_{\RV{M}_1}\left(T\right), &
      p_2 & = 1 - p_1,
    \end{align*}
    with each phase of length $n_1=n\cdot p_1$ and $n_2=n\cdot p_2$, respectively. Each source message is divided into two
    parts of rates $R_\CF$ and $R_\DF$ with the overall rate $R_\DF+R_\CF=R$. Both source transmission parts 
    $\RV{X}_{s,1}$ and $\RV{X}_{s,2}$ are chosen independently and randomly from the 
    sets $\Set{X}_{s,1}$ and $\Set{X}_{s,2}$ with $\left\|\Set{X}_{s,1}\right\|=2^{nR_\CF}$ and $\left\|\Set{X}_{s,2}\right\|=2^{nR_\DF}$.
    Relay node $2$ generates $2^{n\Delta_2}$ quantizations
    $\hat{\RV{Y}}_2$ of length $n_1$ and the same number of broadcast messages $\RV{X}_2$ of length $n_2$. Node $1$
    further creates $2^{nR_{\DF}}$ support messages $\RV{X}_1$ of length $n_1$ at rate $\nicefrac{n}{n_1}R_\DF$.

    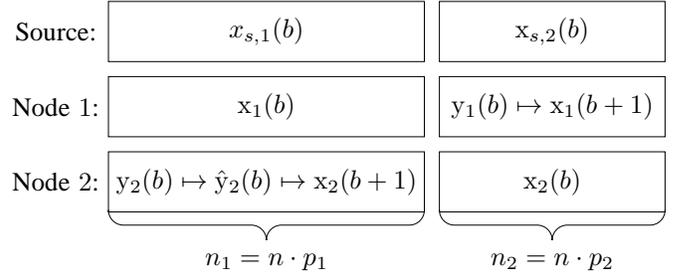
\begin{figure}
      \centering
      \begingroup
\unitlength=1mm
\begin{picture}(87, 36)(0, 0)
  \psset{xunit=1mm, yunit=1mm, linewidth=0.1mm}%
  \psset{arrowsize=3pt 4, arrowlength=1.4, arrowinset=.4}%
  \rput[c](6, 12){Node $2$:}%
  \rput(13, 8)%
  {%
    \psframe(0, 0)(42, 8)
    \rput[c](21, 4){$\RV{y}_2(b)\mapsto\hat{\RV{y}}_2(b)\mapsto\RV{x}_2(b+1)$}
    \psbrace[rot=90, linewidth=0.1mm, ref=t, nodesepB=-2mm](0, 0)(42, 0){$n_1 = n\cdot p_1$}
    \psframe(44, 0)(74, 8)
    \rput[c](59, 4){$\RV{x}_2(b)$}
    \psbrace[rot=90, linewidth=0.1mm, ref=t, nodesepB=-2mm](44, 0)(74, 0){$n_2 = n\cdot p_2$}
  }%
  \rput[c](6, 22){Node $1$:}%
  \rput(13, 18)%
  {%
    \psframe(0, 0)(42, 8)
    \rput[c](21, 4){$\RV{x}_1(b)$}
    \psframe(44, 0)(74, 8)
    \rput[c](59, 4){$\RV{y}_1(b)\mapsto\RV{x}_1(b+1)$}
  }%
  \rput[c](6, 32){Source:}%
  \rput(13, 28)%
  {%
    \psframe(0, 0)(42, 8)
    \rput[c](21, 4){$x_{s,1}(b)$}
    \psframe(44, 0)(74, 8)
    \rput[c](59, 4){$\RV{x}_{s,2}(b)$}
  }%
\end{picture}
\endgroup
      \caption{Coding structure for the mixed strategy with $N=2$ where both nodes are alternately transmitting.}
      \label{figure:halfduplex.random:encoding_mixed_yarp}
    \end{figure}
    Now consider the coding procedure illustrated in Fig. \ref{figure:halfduplex.random:encoding_mixed_yarp}.
    Node $2$ tries to find at the end of phase $1$ in block $b$ an index $q_{2,b+1}$ such that the corresponding
    quantization $\hat{\RV{Y}}_2$ is jointly typical with the node's channel output. In the second phase of block
    $b+1$ node $2$ then transmits the broadcast message assigned to index $q_{2,b+1}$ (there is no advantage in terms
    of achievable rates if node $2$ already transmits the corresponding message in block $b$). 
    Node $1$ decodes at the end of phase $2$ in block $b$ the quantization index of node $2$ by
    taking into consideration that it contains information about the support message of node $1$. Alternatively,
    if the inter-relay channel is rather poor it might skip this step and consider this transmission as interference.
    Afterwards, it decodes the source message $\RV{X}_{s,2}$ and the corresponding
    message index $q_{s,2,b}$. In block $b+1$ the first relay transmits the supporting message
    $\RV{X}_1$ assigned to index $q_{1,b+1}=q_{s,2,b}$.
    
    Obviously, the quantization of node $2$ does not only contain information about the source transmission but also
    about the support information transmitted by node $1$. Our approach exploits this fact as follows: At the end
    of block $b$ the destination decodes at first the quantization of node $2$, i.\,e., $q_{2,b}$. Using this quantization it searches
    for all relay messages jointly typical with this quantization and its own channel output. Then, it reuses
    the quantization decoded at the end of the previous block to search for all source messages jointly
    typical with this quantization and its channel output in block $b-2$. Finally, building the intersection
    of both sets gives the source message index transmitted in the phase $2$ of block $b-2$.
    To decode the message index of the phase $1$ in block $b-2$, it uses the quantization of node $2$
    and its own channel output.

    As mentioned in the previous section, we do not use an intermediate binning of all quantization
    messages to a set of broadcast messages. By decoding both jointly, we avoid the bottleneck of decoding
    at first the broadcast messages and then the quantization separately. Based on the previous
    description we have the following theorem:
    \begin{theorem}\label{theorem:mixed_protocol}
      The previously described mixed protocol achieves any rate $R = R_{\text{DF}} + R_{\text{CF}}$ subject to
      \begin{multline}
	R_{\text{DF}} \leq \sup\limits_p\min\Bigl\{
	p_2\I\left(\RV{X}_{s,2}; \RV{Y}_d | \RV{X}_2\right) + p_1\I\left(\RV{X}_1; \hat{\RV{Y}}_2, \RV{Y}_d | \RV{X}_2\right), \\
	p_2\I\left(\RV{X}_{s,2}; \RV{Y}_1 | \RV{X}_2\right)
	\Bigr\},
	\label{eq:mixed:10}
      \end{multline}
      if node $1$ decodes the quantization of node $2$ and
      \begin{multline}
	R_{\text{DF}} \leq \sup\limits_p\min\Bigl\{
	p_2\I\left(\RV{X}_{s,2}; \RV{Y}_d | \RV{X}_2\right) + p_1\I\left(\RV{X}_1; \hat{\RV{Y}}_2, \RV{Y}_d | \RV{X}_2\right), \\
	p_2\I\left(\RV{X}_{s,2}; \RV{Y}_1\right)
	\Bigr\}
      \end{multline}
      otherwise. Furthermore,
      \begin{equation}
	R_{\text{CF}} \leq \sup\limits_p p_1\I\left(\RV{X}_{s,1}; \hat{\RV{Y}}_2, \RV{Y}_d | \RV{X}_1\right)
	\label{eq:mixed:12}
      \end{equation}
      subject to 
      \begin{equation}
	p_1 \I\left(\hat{\RV{Y}}_2; \RV{Y}_d\right) + p_2 \I\left(\RV{X}_2; \RV{Y}_d\right) \geq p_1 \I\left(\hat{\RV{Y}}_2; \RV{Y}_2\right),
	\label{eq:mixed:14}
      \end{equation}
      and if node $1$ decodes the quantization of node $2$
      \begin{equation}
	p_1 \I\left(\RV{X}_1; \hat{\RV{Y}}_2\right) + p_2 \I\left(\RV{X}_2; \RV{Y}_1\right)  \geq p_1 \I\left(\hat{\RV{Y}}_2; \RV{Y}_2\right).
	\label{eq:mixed:16}
      \end{equation}
      We further have the supremum over all joint pdf of the form
      \begin{equation}
	\begin{split}
	  & p\left(\RV{y}_{[1;3]}, \RV{x}_{s,[1;2]}, \RV{x}_{[1;2]}, \hat{\RV{y}}_2, \RV{m}_{[0;2]}\right) = p\left(\RV{m}_{[0;2]}\right)\\
	  & \quad {\cdot}\; p\left(\RV{y}_{[1;3]} | \RV{x}_{s,[1;2]}, \RV{x}_{[1;2]}, \RV{m}_{[0;2]}\right)\cdot
	    p\left(\hat{\RV{y}}_2 | \RV{y}_2, \RV{m}_{[0;2]}\right) \\
	  & \quad{\cdot}\;\prod\limits_{l=1}^2p\left(\RV{x}_l | \RV{m}_{[0;2]}\right)p\left(\RV{x}_{s,l} | \RV{m}_{[0;2]}\right).
	\end{split}
	\label{eq:mixed:18}
      \end{equation}
    \end{theorem}
    \begin{proof}
      From rate distortion theory we can immediately state that $\Delta_2\geq p_1\I\left(\hat{\RV{Y}}_2; \RV{Y}_2\right)$. Node $1$
      decodes the quantization index at the end of block $b$ iff
      \begin{equation*}
	\begin{split}
	  & \exists \tilde q_{2,b}: \tilde q_{2,b} \in \left\{\hat q_{2,b}: \left(\RV{x}_2\left(\hat q_{2,b}\right), \RV{y}_1(b)\right)\in\StronglyTypicalSet\right\} \\
	  & \quad \wedge \left(\hat{\RV{y}}_2\left(\tilde q_{2,b}\right), \RV{x}_1\left(q_{1,b-1}\right)\right)\in\StronglyTypicalSet,
	\end{split}
      \end{equation*}
      which implies $\Delta_2 \leq p_1\I\left(\RV{X}_1; \hat{\RV{Y}}_2\right) + p_2\I\left(\RV{X}_2; \RV{Y}_1\right)$,
      summarized in (\ref{eq:mixed:16}). Then, node $1$ decodes the source message
      which gives the r.h.s. of the minimum in (\ref{eq:mixed:10}).

      The destination uses the same method as node $1$ to decode the quantization of node $2$ which gives (\ref{eq:mixed:14}).
      To decode the source message at the end of block $b$ it searches for
      \begin{equation*}
	\begin{split}
	  & \exists \tilde q_{s,2,b-2}: \tilde q_{s,2,b-2} = \Bigl\{\hat q_{1,b-1}: \bigl(\RV{x}_1\left(\hat q_{1,b-1}\right), \RV{y}_d\left(b-1\right), \dots \\
	  & \quad \hat{\RV{y}}_2\left(q_{2,b}\right)\bigr)\in\StronglyTypicalSet\Bigr\}
	    \cap \Bigl\{\hat q_{s,2,b-2}: \bigl(\RV{x}_{s,2}\left(\hat q_{s,2,b-2}\right), \dots \\
	  & \quad \hat{\RV{y}}_2\left(q_{2,b-1}\right), \RV{y}_d\left(b-2\right)\bigr)\in\StronglyTypicalSet\Bigr\},
	\end{split}
      \end{equation*}
      which implies the l.h.s. of the minimum in (\ref{eq:mixed:10}). Finally, using
      the quantization message of node $2$ and its own channel output it can
      decode the message transmitted in the first phase which implies the constraint
      given in (\ref{eq:mixed:12}). The proof for achievability again
      follows standard methods \cite{Cover.Thomas.1991}.
    \end{proof}

  \section{Summary and outlook}  
    This paper presented strategies for multiple relay networks 
    constrained by a half-duplex operation. More specifically, we derived achievable rates for
    an $N$-terminal implementation of the decode-and-forward and compress-and-forward approaches as well as
    for a mixed strategy used by two alternately transmitting relay nodes. 
    Based on this paper we will present in our upcoming work achievable rates for wireless channels such as the Gaussian channel.
    \vspace{0.5cm}

  \bibliographystyle{IEEEtran}
  \bibliography{IEEEfull,my-references}

\end{document}